\documentclass[12pt]{article}
\begin{document}
\newcommand{\beq}{\begin{equation}}
\newcommand{\eeq}{\end{equation}}
\newcommand{\beqa}{\begin{eqnarray}}
\newcommand{\eeqa}{\end{eqnarray}}
\newcommand{\beqar}{\begin{eqnarray*}}
\newcommand{\eeqar}{\end{eqnarray*}}
\newcommand{\al}{\alpha}
\newcommand{\be}{\beta}
\newcommand{\del}{\delta}
\newcommand{\D}{\Delta}
\newcommand{\eps}{\epsilon}
\newcommand{\ga}{\gamma}
\newcommand{\Ga}{\Gamma}
\newcommand{\ka}{\kappa}
\newcommand{\inn}{\!\cdot\!}
\newcommand{\h}{\eta}
\newcommand{\kk}{\varphi}
\newcommand\F{{}_3F_2}
\newcommand{\la}{\lambda}
\newcommand{\La}{\Lambda}
\newcommand{\na}{\nabla}
\newcommand{\Om}{\Omega}
\newcommand{\p}{\phi}
\newcommand{\sig}{\sigma}
\renewcommand{\t}{\theta}
\newcommand{\z}{\zeta}
\newcommand{\ssc}{\scriptscriptstyle}
\newcommand{\eg}{{\it e.g.,}\ }
\newcommand{\ie}{{\it i.e.,}\ }
\newcommand{\labell}[1]{\label{#1}} 
\newcommand{\reef}[1]{(\ref{#1})}
\newcommand{\labels}[1]{\vskip-2ex$_{#1}$\label{#1}} 
\newcommand\prt{\partial}
\newcommand\veps{\varepsilon}
\newcommand\ls{\ell_s}
\newcommand\cF{{\cal F}}
\newcommand\cA{{\cal A}}
\newcommand\cS{{\cal S}}
\newcommand\cH{{\cal H}}
\newcommand\cM{{\cal M}}
\newcommand\cL{{\cal L}}
\newcommand\cG{{\cal G}}
\newcommand\cN{{\cal N}}
\newcommand\cl{{\iota}}
\newcommand\cP{{\cal P}}
\newcommand\cV{{\cal V}}
\newcommand\cg{{\it g}}
\newcommand\cR{{\cal R}}
\newcommand\cB{{\cal B}}
\newcommand\cO{{\cal O}}
\newcommand\tcO{{\tilde {{\cal O}}}}
\newcommand\bz{\bar{z}}
\newcommand\bw{\bar{w}}
\newcommand\hF{\hat{F}}
\newcommand\hA{\hat{A}}
\newcommand\hT{\hat{T}}
\newcommand\htau{\hat{\tau}}
\newcommand\hD{\hat{D}}
\newcommand\hf{\hat{f}}
\newcommand\hg{\hat{g}}
\newcommand\hp{\hat{\phi}}
\newcommand\hh{\hat{h}}
\newcommand\ha{\hat{a}}
\newcommand\hQ{\hat{Q}}
\newcommand\hP{\hat{\Phi}}
\newcommand\hb{\hat{b}}
\newcommand\hc{\hat{c}}
\newcommand\hd{\hat{d}}
\newcommand\hS{\hat{S}}
\newcommand\hX{\hat{X}}
\newcommand\tL{\tilde{\cal L}}
\newcommand\hL{\hat{\cal L}}
\newcommand\tG{{\widetilde G}}
\newcommand\tg{{\widetilde g}}
\newcommand\tphi{{\widetilde \phi}}
\newcommand\tPhi{{\widetilde \Phi}}
\newcommand\te{{\tilde e}}
\newcommand\tk{{\tilde k}}
\newcommand\tf{{\tilde f}}
\newcommand\tF{{\widetilde F}}
\newcommand\tK{{\widetilde K}}
\newcommand\tE{{\widetilde E}}
\newcommand\tpsi{{\tilde \psi}}
\newcommand\tX{{\widetilde X}}
\newcommand\tD{{\widetilde D}}
\newcommand\tO{{\widetilde O}}
\newcommand\tS{{\tilde S}}
\newcommand\tB{{\widetilde B}}
\newcommand\tA{{\widetilde A}}
\newcommand\tT{{\widetilde T}}
\newcommand\tC{{\widetilde C}}
\newcommand\tV{{\widetilde V}}
\newcommand\thF{{\widetilde {\hat {F}}}}
\newcommand\Tr{{\rm Tr}}
\newcommand\tr{{\rm tr}}
\newcommand\STr{{\rm STr}}
\newcommand\M[2]{M^{#1}{}_{#2}}
\parskip 0.3cm

\vspace*{1cm}

\begin{center}
{\bf \Large   S-duality invariant dilaton couplings  at order $\alpha'^3$   }

\vspace*{1cm}

{  Mohammad R. Garousi\footnote{garousi@um.ac.ir} }\\
\vspace*{1cm}
{ Department of Physics, Ferdowsi University of Mashhad,\\ P.O. Box 1436, Mashhad, Iran}
\\
\vspace{2cm}

\end{center}

\begin{abstract}
\baselineskip=18pt

The  Riemann curvature correction to the type II supergravity  at eight-derivative level  is given schematically as $(t_8t_8+\frac{1}{8}\eps_{10}\eps_{10})R^4$ at tree-level. The replacement of  the generalized Riemann curvature in $t_8t_8R^4$,   proposed by Gross and Sloan, produces various NS-NS couplings which are invariant under T-duality. Recently, using the combination of S-duality and T-duality transformations on these couplings, we have found groups of couplings which are invariant under the    S-duality transformation.  In this paper, we have examined the   couplings   involving the dilaton  with  direct scattering amplitude calculations of  four NS-NS vertex operators  in the superstring theory and found exact agreement. 

The coupling $\eps_{10}\eps_{10}R^4$ is a total derivative term at four-field level. The $\sigma$-model beta function approach implies the presence of this term at the tree-level. By examining the sphere-level scattering amplitude of five gravitons, we have also confirmed the presence of this term in the tree-level effective action.

\end{abstract}
Keywords: Dilaton coupling, S-duality, S-matrix 

\setcounter{page}{0}
\setcounter{footnote}{0}
\newpage
\section{Introduction  } \label{intro}

One of the nonperturbative symmetries in superstring theory is the S-duality of type IIB superstring theory  \cite{Font:1990gx,Sen:1994fa,Rey:1989xj,Sen:1994yi,Schwarz:1993cr,Hull:1994ys}.
   It relates the type IIB theory at weak (strong) coupling to the type IIB at strong (weak) coupling. At low energy, this is the symmetry of type IIB  supergravity which contains the following bosonic couplings in the Einstein frame \cite{Becker:2007zj}:
\beqa
S\!\supset\!\!\frac{1}{2 }\int d^{10}x  \sqrt{-G}\bigg[R+\frac{1}{4}\Tr(\cM_{,\mu}\cM^{-1}_{,\mu})-\frac{1}{12}\cH^T_{\mu\nu\rho}\cM\cH_{\mu\nu\rho}-\frac{1}{4}|\tF_{(5)}|^2\bigg]\!-\frac{1}{8}\int C_{(4)} \cH^T\cN \cH\labell{super}
\eeqa
where the five-form field strength is $\tF_{(5)}=dC_{(4)}+\frac{1}{2}\cB^T\cN\cH$, and the self-duality condition  $\tF_{(5)}=*\tF_{(5)}$ is imposed by hand\footnote{We  use only subscript indices and the repeated indices are contracted with the  inverse of  metric.  Our conventions also set   the gravitational coupling constant $\kappa=1$.}. The metric and the R-R four-form are invariant under the $SL(2,R)$ transformations. The B-field and the R-R two-form transform as doublet  \cite{ Tseytlin:1996it,Green:1996qg}, \ie
\beqa
\cB\equiv\pmatrix{B \cr 
C_{(2)}}\rightarrow (\Lambda^{-1})^T \pmatrix{B \cr 
C_{(2)}}\,\,\,;\,\,\,\Lambda=\pmatrix{p&q \cr 
r&s}\in SL(2,R)\labell{2}
\eeqa
Since the parameters of the $SL(2,R)$ transformations are constant, their field strengths $\cH=d\cB$    also transform as doublet. 
 The dilaton and the R-R scalar transform nonlinearly as $\tau\rightarrow \frac{p\tau+q}{r\tau+s}$ where the complex scalar field is defined as $\tau=C_{(0)}+ie^{-\phi}$. The matrix $\cM$ defined in terms of the dilaton and the R-R scalar, \ie 
 \beqa
 {\cal M}=e^{\phi}\pmatrix{|\tau|^2&C_{(0)} \cr 
C_{(0)}&1}\labell{M}
\eeqa
then  transforms  as \cite{Gibbons:1995ap}
\beqa
{\cal M}\rightarrow \Lambda {\cal M}\Lambda ^T\labell{M1}
\eeqa
The derivatives of the matrix $\cM$ also transform as \reef{M1}. The matrix $\cN$ which is defined as 
 \beqa
 {\cal N}=\pmatrix{0&1 \cr 
-1&0}\labell{N}
\eeqa
  can be written as $
\Lambda {\cal N}\Lambda ^T$. Using these matrices and the transformation \reef{2}, one observes that the  supergravity action \reef{super} is invariant under the   $SL(2,R)$ transformations.
 The stringy behaviors of the superstring theory which are encoded in both higher derivative corrections and the genus corrections, however,  are not  captured by the above classical action.  Moreover, the $SL(2,R)$ transformations of the classical theory are broken to the $SL(2,Z)$ transformations. Therefore, the stringy   corrections   must be invariant under the $SL(2,Z)$ transformations.  

The   higher derivative corrections to the supergravity \reef{super} start at the eight-derivative level, and were first found   from the sphere-level four-graviton scattering amplitude   \cite{ Schwarz:1982jn,Gross:1986iv} as well as from the $\sigma$-model beta function approach \cite{Grisaru:1986vi,Freeman:1986zh}. The result in the Einstein frame is  
\beqa
S\supset \frac{\gamma \z(3)}{3.2^7} \int d^{10}x e^{-3\phi/2} \sqrt{-G}(t_8t_8R^4+\frac{1}{8}\eps_{10}\eps_{10}R^4)\labell{Y1}
\eeqa
where $\gamma=\frac{\alpha'^3}{2^{5}}$ and we have used  the normalization of the tensor $t_8$ as in \reef{t8}.  
The couplings given by $t_8t_8R^4$ have nonzero contribution at four-graviton level, so they were found from the sphere-level S-matrix element of four graviton vertex operators \cite{ Schwarz:1982jn,Gross:1986iv}, whereas the couplings given by $\eps_{10}\eps_{10}R^4$ have nonzero contribution at five-graviton level \cite{Zumino:1985dp} which have not yet been confirmed by the sphere-level S-matrix element of five graviton vertex operators. However, the presence of this term in the tree-level effective action was dictated by the $\sigma$-model beta function approach \cite{Grisaru:1986vi,Freeman:1986zh}. In this paper, among other things, we will show that the sphere-level scattering amplitude of five gravitons confirms the presence of $\eps_{10}\eps_{10}R^4$ in the tree-level effective action.

Unlike the two-derivative level  supergravity action \reef{super} which is invariant under the S-duality without adding loop or nonperturbative effects, the presence of the dilaton factor in the eight-derivative level      indicates that the action \reef{Y1} needs loop and nonperturbative corrections to become S-duality invariant. The   $SL(2,Z)$ invariant form of the action \reef{Y1} has been conjectured in  \cite{Green:1997tv} - \cite{Basu:2007ck} to be
\beqa
S\supset\frac{\gamma }{3.2^8 }\int d^{10}x E_{(3/2)}(\tau,\bar{\tau}) \sqrt{-G}(t_8t_8R^4+\frac{1}{8}\eps_{10}\eps_{10}R^4)\labell{Y2}
\eeqa
where $E_{(3/2)}(\tau,\bar{\tau}) $ is the $SL(2,Z)$ invariant non-holomorphic Eisenstein series which has the following weak-expansion \cite{Green:1997tv}:
\beqa
E_{(3/2)}(\tau,\bar{\tau})
&\!\!\!\!=\!\!\!\!&2\z(3)\tau_2^{3/2}+4\z(2)\tau_2^{-1/2}+8\pi\tau_2^{1/2}\sum_{m\neq 0,n\geq 1}\left|\frac{m}{n}\right| K_{1}(2\pi|mn|\tau_2)e^{2\pi imn\tau_1}\labell{series}
\eeqa
where $\tau=\tau_1+i\tau_2 $ and $K_1$ is the Bessel function. The Eisenstein series which appear in various 16 fermion couplings are completely determined by supersymmetry and the structure of the moduli space \cite{Green:1998by}. Linear supersymmetry transformations  then relates those couplings to the $R^4$  couplings  \cite{Green:1998by}. The above expansion shows that there are no perturbative corrections beyond the tree level and one-loop level, but there are an infinite number  of D-instanton  corrections.  By explicit calculation, it has been shown in \cite{D'Hoker:2005jc} that there is no two-loop correction to the action \reef{Y2}. The coupling $\eps_{10}\eps_{10}R^4$ at one-loop level has been confirmed in \cite{Richards:2008jg,Liu:2013dna} by explicit calculation of  torus-level S-matrix element of  five graviton vertex operators.  

 The B-field and dilaton couplings have been added to $t_8t_8R^4$     by extending the Riemann curvature to the generalized Riemann curvature \cite{Gross:1986mw}\footnote{Note that the normalizations of the dilation  and B-field here  are $\sqrt{2}$ and 2 times the normalization of the dilaton and B-field in \cite{Gross:1986mw}, respectively.} 
 \beqa
 \bar{R}_{ab}{}^{cd}&= &R_{ab}{}^{cd}- \eta_{[a}{}^{[c}\phi_{,b]}{}^{d]}+  e^{-\phi/2}H_{ab}{}^{[c,d]}\labell{trans}
\eeqa
where   the bracket notation is $H_{ab}{}^{[c,d]}=\frac{1}{2}(H_{ab}{}^{c,d}-H_{ab}{}^{d,c})$, and comma denotes the partial derivative. The resulting couplings are then invariant under T-duality \cite{Garousi:2012jp,Garousi:2013zca,Liu:2013dna}. Imposing the invariance of these couplings under  the combination of S-duality and linear T-duality, some new S-duality invariant actions at eight-derivative level  have been found in \cite{Garousi:2013lja}.  The S-duality invariant action which includes, among other things,  the couplings of two gravitons and two dilatons is   \cite{Garousi:2013lja}
 \beqa
S &\supset&\frac{\gamma}{2 } \int d^{10}x E_{(3/2)}(\tau,\bar{\tau})\sqrt{-G} \bigg[-\frac{1}{16}\Tr[\cM_{,qr}\cM^{-1}_{,qr} ] R^2_{h k mn}-\frac{1}{2} \Tr[\cM_{,ph}\cM^{-1}_{,pr} ] R_{h k m n}R_{mnkr} \nonumber\\&&-\frac{1}{2}  \Tr[\cM_{,mr}\cM^{-1}_{,ph} ] R_{h k m n}R_{npkr} -\frac{1}{2}  \Tr[\cM_{,mh}\cM^{-1}_{,pr} ] R_{h k m n}R_{npkr}\bigg]\labell{F1R2}
\eeqa
 The S-duality invariant action which includes, among other things, the couplings of two B-fields and two dilatons is \cite{Garousi:2013lja}\footnote{Note that the normalization of the Eisenstein series and the normalization of the form field strengths here are different from the corresponding terms in \cite{Garousi:2013lja}, \ie $E_{(3/2)}|_{\rm here}=2\z(3)E_{3/2}|_{\rm there}$, $H|_{\rm here}=dB$ and  $H|_{\rm there}=\frac{1}{2}dB$, $F|_{\rm here}=dC$ and  $F|_{\rm there}=\frac{1}{2}dC$.}
 \beqa
S &\supset&\frac{\gamma}{8 } \int d^{10}x E_{(3/2)}(\tau,\bar{\tau})\sqrt{-G} \bigg[\frac{1}{6} \Tr[\cM_{,nh}\cM^{-1}_{,nk} ] \cH^T_{m p r,h}\cM \cH_{m p r,k}\labell{F1HS}\\&&-\frac{1}{2} \Tr[\cM_{,nh}\cM^{-1}_{,km} ] \cH^T_{m p r,k} \cM\cH_{n p r,h}-\frac{1}{2} \Tr[\cM_{,hm}\cM^{-1}_{,kn} ] \cH^T_{m p r,k} \cM\cH_{n p r,h}\bigg]\nonumber
\eeqa
  The S-duality invariant action which includes, among other things,  the couplings of one dilaton, one graviton and two B-fields  is \cite{Garousi:2013lja}
\beqa
S &\supset&\frac{\gamma}{8 } \int d^{10}x E_{(3/2)}(\tau,\bar{\tau})\sqrt{-G}\bigg[ 4 \cH^T_{h q r,n} \cM_{,rm}\cH_{k n p,h} R_{m p k q}-4   \cH^T_{n p r,h}\cM_{,mk} \cH_{h n q,r} R_{k q m p}\nonumber\\&&-4   \cH^T_{n p q,m} \cM_{,qh}\cH_{m p r,k} R_{k r h n}+4   \cH^T_{n p q,h} \cM_{,qm}\cH_{k n r,h} R_{m r k p}-2   \cH^T_{n p q,h} \cM_{,mh}\cH_{n p r,k} R_{m r k q}\nonumber\\&&+2 \cH^T_{m n q,h} \cM_{,rk} \cH_{n p q,k} R_{p r h m}-2 \cH^T_{m n p,k} \cM_{,rm} \cH_{n p q,h} R_{q r h k}\bigg]\labell{F1F3}
\eeqa

The   S-duality invariant action   containing four dilatons can also be found from the couplings of four $H$s in $t_8t_8\bar{R}^4$.  The $SL(2,R)$ invariant form of the coupling $t_8t_8(\partial H)^4$ requires the coupling $t_8t_8(\partial F)^4$ for the R-R three-form field strengths.  Then following  the linear T-duality prescription given in \cite{Garousi:2013lja},   one can find the couplings of four   R-R one-form field strengths which is $3.2^5(F_{n,m}F_{n,m})^2$. Then using the following  expression:
\beqa
 \Tr[\cM_{,hk}\cM^{-1}_{,mn} ]&=&-2e^{2\phi}F_{h,k}F_{m,n}-2\phi_{,hk}\phi_{,mn}\labell{FF}
\eeqa 
one may find the S-duality invariant coupling
 \beqa
S \supset\frac{\gamma}{32 } \int d^{10}x E_{(3/2)}(\tau,\bar{\tau})\sqrt{-G} \bigg[ a( \Tr[\cM_{,nm}\cM^{-1}_{,nm} ])^2 +b \Tr[\cM_{,nm}\cM^{-1}_{,hk} ]\Tr[\cM_{,hk}\cM^{-1}_{,nm} ]\bigg]\labell{F1HS1}
\eeqa
where the constants $a$ and $b$ satisfy $a+b=1$. The combination of the dilaton couplings in \reef{F1HS1} and \reef{F1R2}, and the four graviton couplings in \reef{Y2} should give the coupling $t_8t_8\hat{R}^4$ where $\hat{R}$ is the first two terms on the right hand side of \reef{trans}. In the string frame, the dilaton term in \reef{trans} disappears \cite{Garousi:2012jp}, so the Einstein frame  coupling $t_8t_8\hat{R}^4$  transforms to $t_8t_8R^4$ in the string frame. Hence, there is no dilaton coupling in the string frame.

In this paper, we are going to confirm the   actions \reef{F1R2}, \reef{F1HS}, \reef{F1F3} and \reef{F1HS1} by comparing the four NS-NS couplings in them with the direct calculation of the S-matrix element of four NS-NS vertex  operators in the type IIB superstring theory.

The outline of the paper is as follows: We begin in section 2 by reviewing the tree level and one-loop level scattering amplitude of four NS-NS vertex operators in  the type IIB superstring theory. In subsection 2.1, we show that the couplings of two dilatons and two gravitons extracted from the S-duality invariant action \reef{F1R2} are exactly reproduced by the corresponding superstring theory S-matrix element. In subsections 2.2, 2.3, and 2.4, we show that the dilaton couplings in actions \reef{F1HS1}, \reef{F1HS}, and \reef{F1F3}, respectively, are reproduced by the appropriate S-matrix elements in the superstring theory. In section 3, using the observation that the sphere-level S-matrix element in RNS formalism must contain terms which are proportional to at least two Mandelstam variables \cite{Garousi:2012yr,Barreiro:2012aw}, we show that the massless poles and the contact terms of five-graviton amplitude in the field theory calculated from the Hilbert-Einstein term in \reef{super} and from the couplings $t_8t_8R^4+\frac{\alpha}{2}\eps_{10}\eps_{10}R^4$, satisfy this condition when the constant $\alpha=\frac{1}{4}$.  This confirms the tree-level action \reef{Y1}.
 
\section{Four-point amplitude}

The sphere-level scattering amplitude of four massless NS-NS vertex operators  with polarization tensors $\veps^{ab}$ in covariant formalism is given by \cite{Schwarz:1982jn}
\beqa
{\cal A} =-  i\frac{\gamma}{2}e^{-2\phi}\frac{\Gamma(-s/8)\Gamma(-t/8)\Gamma(-u/8)}{\Gamma(1+s/8)\Gamma(1+t/8)\Gamma(1+u/8)}\veps_1^{a_1b_1}\veps_2^{a_2b_2}\veps_3^{a_3b_3}\veps_4^{a_4b_4}K_{a_1a_2a_3a_4}K_{b_1b_2b_3b_4}\labell{amp1}
\eeqa 
There is also a factor of delta function $\delta^{10}(k_1+k_2+k_3+k_4)$ imposing conservation of momentum. The Mandelstam variables $s=-4\alpha'k_1\inn k_2$, $t=-4\alpha'k_1\inn k_3$, $u=-4\alpha'k_2\inn k_3$ satisfy $s+t+u=0$, and 
\beqa
K_{a_1a_2a_3a_4}&\!\!\!\!\!=\!\!\!\!\!&4\bigg[- k_2.k_1 k_3.k_1 \eta _{{a1a4}} \eta _{{a2a3}}- k_2.k_1 k_2.k_3 \eta _{{a1a3}} \eta _{{a2a4}}- k_2.k_3 k_3.k_1 \eta _{{a1a2}} \eta _{{a3a4}}\nonumber\\&&+ k_2.k_1 \eta _{{a1a4}} \left(k_1\right)_{{a2}} \left(k_1\right)_{{a3}}+ k_3.k_1 \eta _{{a1a4}} \left(k_1\right)_{{a2}} \left(k_1\right)_{{a3}}- k_2.k_1 \eta _{{a2a4}} \left(k_1\right)_{{a3}} \left(k_2\right)_{{a1}}\nonumber\\&&- k_3.k_1 \eta _{{a2a4}} \left(k_1\right)_{{a3}} \left(k_2\right)_{{a1}}+ k_3.k_1 \eta _{{a1a4}} \left(k_1\right)_{{a2}} \left(k_2\right)_{{a3}}- k_3.k_1 \eta _{{a2a4}} \left(k_2\right)_{{a1}} \left(k_2\right)_{{a3}}\nonumber\\&&+ k_2.k_1 \eta _{{a1a2}} \left(k_1\right)_{{a3}} \left(k_2\right)_{{a4}}+ k_3.k_1 \eta _{{a1a2}} \left(k_1\right)_{{a3}} \left(k_2\right)_{{a4}}+ k_3.k_1 \eta _{{a1a2}} \left(k_2\right)_{{a3}} \left(k_2\right)_{{a4}}\nonumber\\&&- k_2.k_1 \eta _{{a3a4}} \left(k_1\right)_{{a2}} \left(k_3\right)_{{a1}}- k_3.k_1 \eta _{{a3a4}} \left(k_1\right)_{{a2}} \left(k_3\right)_{{a1}}+ k_2.k_1 \eta _{{a2a4}} \left(k_2\right)_{{a3}} \left(k_3\right)_{{a1}}\nonumber\\&&- k_2.k_1 \eta _{{a2a3}} \left(k_2\right)_{{a4}} \left(k_3\right)_{{a1}}+ k_2.k_1 \eta _{{a1a4}} \left(k_1\right)_{{a3}} \left(k_3\right)_{{a2}}+ k_3.k_1 \eta _{{a3a4}} \left(k_2\right)_{{a1}} \left(k_3\right)_{{a2}}\nonumber\\&&+ k_2.k_1 \eta _{{a1a3}} \left(k_2\right)_{{a4}} \left(k_3\right)_{{a2}}- k_2.k_1 \eta _{{a3a4}} \left(k_3\right)_{{a1}} \left(k_3\right)_{{a2}}+ k_2.k_1 \eta _{{a1a3}} \left(k_1\right)_{{a2}} \left(k_3\right)_{{a4}}\nonumber\\&&+ k_3.k_1 \eta _{{a1a3}} \left(k_1\right)_{{a2}} \left(k_3\right)_{{a4}}- k_3.k_1 \eta _{{a2a3}} \left(k_2\right)_{{a1}} \left(k_3\right)_{{a4}}+ k_3.k_1 \eta _{{a1a2}} \left(k_2\right)_{{a3}} \left(k_3\right)_{{a4}}\nonumber\\&&+ k_2.k_1 \eta _{{a1a3}} \left(k_3\right)_{{a2}} \left(k_3\right)_{{a4}}\bigg]\labell{kin1}
\eeqa
The on-shell conditions are $k_i^2=k_i\inn\veps_i=\veps_i\inn k_i=0$. The polarization tensor is symmetric and traceless for graviton, antisymmetric for B-field and for dilaton it is 
\beqa
\veps^{ab}= \frac{\phi}{\sqrt{8}}(\eta^{ab}-k^a\ell^b-k^b\ell^a)\labell{del}
\eeqa
where $\ell^a$ is an auxiliary vector which satisfies $k\inn\ell=1$ and $\phi$ is the dilaton polarization which is one. In  equation \reef{kin1} we have used the conservation of momentum   to write the amplitude in terms of momentum $k_1,k_2,k_3$. We have also used the on-shell conditions to rewrite $k_1\inn\veps_4=-k_2\inn\veps_4-k_3\inn\veps_4$, similarly for $\veps_4\cdot k_1$. We have normalized the amplitude \reef{amp1} to be consistent with the normalization factor in the  couplings \reef{Y1}. The factor of $i$ corresponds to that of the analogous field theory amplitudes calculated in Minkowski space.

The tree-level coupling $t_8t_8R^4$  has been found in \cite{Gross:1986iv}  from  the amplitude \reef{amp1} by expanding  the gamma functions at low energy  
\beqa
\frac{\Gamma(-s/8)\Gamma(-t/8)\Gamma(-u/8)}{\Gamma(1+s/8)\Gamma(1+t/8)\Gamma(1+u/8)}&=&-\frac{2^9}{stu}-2\z(3)+\cdots
\eeqa
The first term corresponds to the massless poles in the four-point function which are reproduced by the Einstein-Hilbert action \cite{Sannan:1986tz}, and the second term,
\beqa
\Delta{\cal A}&=& i\gamma\z(3)e^{-2\phi} \veps_1^{a_1b_1}\veps_2^{a_2b_2}\veps_3^{a_3b_3}\veps_4^{a_4b_4}K_{a_1a_2a_3a_4}K_{b_1b_2b_3b_4} \labell{kin2}
\eeqa
is the  tree-level $\alpha'^3$-correction corresponding to four NS-NS states.  

The torus-level scattering amplitude  of four massless NS-NS vertex operators has been calculated in \cite{Green:1981xx}, \ie
\beqa
{\cal A}^{\rm 1-loop} &= & i\gamma\pi I(s,t,u) \veps_1^{a_1b_1}\veps_2^{a_2b_2}\veps_3^{a_3b_3}\veps_4^{a_4b_4}K_{a_1a_2a_3a_4}K_{b_1b_2b_3b_4}\labell{amp11}
\eeqa 
where $I(s,t,u)$ is a function which includes massive poles,  massive double poles and threshold branch cuts \cite{D'Hoker:1993ge,D'Hoker:1993mr, D'Hoker:1994yr}. It has the following low energy expansion \cite{Green:1999pv,Richards:2008jg}:
\beqa
I(s,t,u)&=&\frac{\pi}{3}+O(\alpha')
\eeqa
which results the following one-loop correction at order $\alpha'^3$:
\beqa
\Delta{\cal A}^{\rm 1-loop}&=& 2i\gamma\z(2) \veps_1^{a_1b_1}\veps_2^{a_2b_2}\veps_3^{a_3b_3}\veps_4^{a_4b_4}K_{a_1a_2a_3a_4}K_{b_1b_2b_3b_4} \labell{kin21}
\eeqa
The amplitudes \reef{kin2} and \reef{kin21} are in the string frame, \ie  $G^s_{ab}=e^{\phi/2}G_{ab}$. Using the fact that $\sqrt{-G}t_8t_8R^4$ in \reef{Y2} transforms to $e^{-\phi/2}\sqrt{-G^s}t_8t_8R^4$ in the string frame, one observes that the dilaton factors as well as the factors $\z(3)$ and $\z(2)$ in the Eisenstein series in \reef{Y2} are consistent with the amplitudes \reef{kin2} and \reef{kin21}. As a result, if the dilaton actions \reef{F1R2}, \reef{F1HS}, \reef{F1F3} and \reef{F1HS1} at tree-level are consistent with sphere-level scattering amplitude \reef{kin2}, then these couplings at one-loop level would be consistent with the torus-level amplitude \reef{kin21}. So in the following subsections, we compare these actions with the amplitude \reef{kin2}.

 \subsection{Two-dilaton-two-graviton amplitude}
 
It has been shown in \cite{Garousi:2012yr} that the  amplitude \reef{kin2} is zero for one dilaton and three gravitons, and for three dilatons and one graviton. They are consistent with the fact that it is impossible to construct $SL(2,R)$ invariant couplings which have such components. However, the amplitude \reef{kin2} is non-zero for two dilatons and two gravitons. The result   is \cite{Garousi:2012yr} 
\beqa
\Delta{\cal A}  &=& i\frac{\gamma\z(3)}{2}e^{-2\phi}\bigg[ 16 \left(k_2.k_3\right){}^2 \left(k_3.k_1\right){}^2   \Tr\left[\veps _3.\veps _4\right]+ 16 \left(k_2.k_1\right){}^2 k_1.\veps _3.k_1 k_2.\veps _4.k_2  \nonumber\\
&&+32 k_2.k_1 k_3.k_1 k_1.\veps _3.k_1 k_2.\veps _4.k_2  +16 \left(k_3.k_1\right){}^2 k_1.\veps _3.k_1 k_2.\veps _4.k_2  \nonumber\\
&&+32 k_2.k_1 k_3.k_1 k_1.\veps _3.k_2 k_2.\veps _4.k_2  +32 \left(k_3.k_1\right){}^2 k_1.\veps _3.k_2 k_2.\veps _4.k_2  \nonumber\\
&&+16 \left(k_3.k_1\right){}^2 k_2.\veps _3.k_2 k_2.\veps _4.k_2  +32 k_2.k_1 k_3.k_1 k_1.\veps _3.k_2 k_2.\veps _4.k_3  \nonumber\\
&&+32 \left(k_3.k_1\right){}^2 k_1.\veps _3.k_2 k_2.\veps _4.k_3 +32 \left(k_3.k_1\right){}^2 k_2.\veps _3.k_2 k_2.\veps _4.k_3  \nonumber\\
&&+16 \left(k_3.k_1\right){}^2 k_2.\veps _3.k_2 k_3.\veps _4.k_3  +32 \left(k_2.k_1\right){}^2 k_3.k_1 k_1.\veps _3.\veps _4.k_2  \nonumber\\
&&+64 k_2.k_1 \left(k_3.k_1\right){}^2 k_1.\veps _3.\veps _4.k_2  +32 \left(k_3.k_1\right){}^3 k_1.\veps _3.\veps _4.k_2  \nonumber\\
&&+32 k_2.k_1 \left(k_3.k_1\right){}^2 k_2.\veps _3.\veps _4.k_2  +32 \left(k_3.k_1\right){}^3 k_2.\veps _3.\veps _4.k_2  \nonumber\\
&&+32 k_2.k_1 \left(k_3.k_1\right){}^2 k_2.\veps _3.\veps _4.k_3  +32 \left(k_3.k_1\right){}^3 k_2.\veps _3.\veps _4.k_3  \bigg]\phi_1\phi_2\labell{amp2}  
\eeqa
where $\phi_1$ and $\phi_2$ are the polarizations of the dilatons which are one. However, for clarity we keep them. We have also divided the amplitude in \cite{Garousi:2012yr} by $2$ because the normalization of the dilaton in sec.1  is $\sqrt{2}$ times the  normalization of the dilaton vertex operator. 
 
 Now we have to compare the above S-matrix element with the corresponding couplings in \reef{F1R2}. Using the expression \reef{FF}, 
one finds the following couplings of two dilatons and two Riemann curvatures at tree-level in the Einstein frame: 
 \beqa
S &\supset&\frac{\gamma\z(3)}{2 } \int d^{10}x\sqrt{-G}e^{-3\phi/2} \bigg[\frac{1}{4} \phi_{,qr}^2 R_{h k m n}^2+2 \phi_{,ph} \phi_{,pr} R_{h k m n} R_{m n k r}\labell{F1R1}\\
&&\qquad\qquad\qquad\qquad\qquad\quad+2 \phi_{,mr} \phi_{,ph} R_{h k m n} R_{n p k r}+2 \phi_{,mh} \phi_{,pr} R_{h k m n} R_{n p k r}\bigg]\nonumber
\eeqa
Using the perturbation $G_{ab}=\eta_{ab}+2h_{ab}$ where $h_{ab}$ is the graviton, the Riemann curvature at the linear order of the graviton becomes
\beqa
R_{abcd}&=& h_{ad,bc}+h_{bc,ad}-h_{ac,bd}-h_{bd,ac}\labell{R}
\eeqa
Transforming the couplings \reef{F1R1} to the string frame, one finds the overall dilaton factor   to be $e^{-2\phi}$ as in the string amplitude \reef{amp2}. Using conservation of momentum and on-shell relations     to write the momentum space couplings  in terms of the momentum $k_1,k_2,k_3$ and   to rewrite $k_1\inn\veps_4=-k_2\inn\veps_4-k_3\inn\veps_4$, we have checked explicitly that the   couplings of two dilatons and two gravitons in \reef{F1R1} are exactly the same as the couplings in \reef{amp2}.

 \subsection{Four-dilaton  amplitude}
 
The amplitude \reef{kin2} for four dilatons has been found in \cite{Garousi:2012yr} to be
\beqa
\Delta{\cal A} &=&i \frac{\gamma\z(3)}{4}e^{-2\phi}\bigg[ 16 \left(\left(k_2.k_1\right){}^2+k_2.k_1 k_3.k_1+\left(k_3.k_1\right){}^2\right){}^2  
\bigg] \phi_1\phi_2\phi_3\phi_4  \labell{extra2}
\eeqa
where we have also divided the amplitude in \cite{Garousi:2012yr} by $4$ because the normalization of the dilaton in sec.1  is $\sqrt{2}$ times the  normalization of the dilaton vertex operator.  

On the other hand, using the expression \reef{FF}, one finds the following coupling of four dilatons at tree-level in \reef{F1HS1}:
\beqa
S &\supset&\frac{\gamma\z(3)}{4 } \int d^{10}x e^{-3\phi/2} \sqrt{-G} \bigg[  \phi_{,nh}\phi_{,nh}  \bigg]^2 
\eeqa
In the string frame, one finds the overall dilaton factor   to be $e^{-2\phi}$ as in the string amplitude. Transforming the above coupling to the momentum space, \ie labeling the dilatons by the particle labels 1,2,3,4 and adding 24 permutations, one finds exactly the momentum space couplings \reef{extra2} after using conservation of momentum and the on-shell relation $k_i\inn k_i=0$. This confirms the constants $a$ and $b$  in \reef{F1HS1} to satisfy $a=1-b$. The constant $b$, on the other hand,  can be fixed by comparing the couplings of two dilatons and two R-R one-form field strengths in \reef{F1HS1} with the corresponding scattering amplitude in string theory \cite{BG}. 

 \subsection{Two-dilaton-two-B-field amplitude}

The scattering amplitude \reef{kin2} for  two dilatons   and two B-fields has been found in \cite{Garousi:2012yr} to be  
\beqa
&\!\!\!\!\!\!\!\!\!\!\!\!&\Delta{\cal A}=i\frac{\gamma \z(3)}{8}e^{-2\phi} \bigg[2k_3.k_1 \left(2 \left(k_1.\epsilon _3.\epsilon _4.k_2+k_2.\epsilon _3.\epsilon _4.k_2+k_2.\epsilon _3.\epsilon _4.k_3\right)+k_3.k_1  {\Tr}\left[\epsilon _3.\epsilon _4\right]\right)  \labell{extra5}\\
&\!\!\!\!\!\!\!\!\!\!\!\!& +(16 k_3.k_1 k_3.k_2 -8( k_2.k_1 )^2)\left(-2 k_1.\epsilon _3.k_2 k_2.\epsilon _4.k_3+k_2.k_1 \left(2 k_1.\epsilon _3.\epsilon _4.k_2+k_3.k_1  {\Tr}\left[\epsilon _3.\epsilon _4\right]\right)\right)  \bigg]\phi_1\phi_2\nonumber
\eeqa
where we have also divided the amplitude in \cite{Garousi:2012yr} by $8$ because the normalization of the B-field in sec.1  is twice  the  normalization of the B-field vertex operator, and the normalization of the dilaton in sec.1  is $\sqrt{2}$ times the  normalization of the dilaton vertex operator. 

The above amplitude should be compared with the corresponding couplings in \reef{F1HS}. 
Using the expression \reef{FF}, and
\beqa
 \cH^T\cM\cH =e^{-\phi} HH+\cdots\labell{SHH}
\eeqa
where dots refer to the other terms which involve the R-R fields, one finds the following couplings of two dialtons and two B-fields:
\beqa
S &\supset&\frac{\gamma\z(3)}{8 } \int d^{10}x\sqrt{-G}e^{-5\phi/2} \bigg[-\frac{2}{3} \phi_{,nh} \phi_{,nk} H_{m p r,h} H_{m p r,k}+2 \phi_{,hn} \phi_{,km} H_{m p r,k} H_{n p r,h}\nonumber\\
&&\qquad\qquad\qquad\qquad\qquad\quad+2 \phi_{,hm} \phi_{,kn} H_{m p r,k} H_{n p r,h}\bigg]
\eeqa
The B-field strength is $H_{abc}=B_{ab,c}+B_{ca,b}+B_{bc,a}$. In the string frame, one again finds the overall dilaton factor   to be $e^{-2\phi}$.  Using conservation of momentum and on-shell relations, we have checked explicitly that the momentum space couplings of two dilatons and two B-fields in above equation are exactly the same as the couplings in \reef{extra5}.

\subsection{One-dilaton-one-graviton-two-B-field amplitude}

The scattering amplitude \reef{kin2} for one dilaton, one graviton and two B-fields can be calculated by inserting one of the polarization tensors by \reef{del},   one by symmetric and traceless polarization and the other two by antisymmetric polarization tensors. We have found that the auxiliary vector $\ell$ is chancel in the amplitude. The result is 
\beqa
\Delta{\cal A}&=&i\frac{ \gamma \z(3)}{8} e^{-2\phi} \bigg[ -32 \left(k_2.k_1\right){}^2 k_2.\epsilon _4.k_3 k_1.\epsilon _2.\epsilon _3.k_1-32 k_2.k_1 k_3.k_1 k_2.\epsilon _4.k_3 k_1.\epsilon _2.\epsilon _3.k_1\nonumber\\&&-32 k_2.k_1 k_3.k_1 k_2.\epsilon _4.k_3 k_1.\epsilon _2.\epsilon _3.k_2-32 \left(k_2.k_1\right){}^2 k_1.\epsilon _3.k_2 k_1.\epsilon _2.\epsilon _4.k_3\nonumber\\&&-32 k_2.k_1 k_3.k_1 k_1.\epsilon _3.k_2 k_1.\epsilon _2.\epsilon _4.k_3+32 \left(k_2.k_1\right){}^2 k_1.\epsilon _2.k_3 k_1.\epsilon _3.\epsilon _4.k_2\nonumber\\&&+32 k_2.k_1 k_3.k_1 k_1.\epsilon _2.k_3 k_1.\epsilon _3.\epsilon _4.k_2+32 \left(k_2.k_1\right){}^2 k_3.\epsilon _2.k_3 k_1.\epsilon _3.\epsilon _4.k_2\nonumber\\&&+32 \left(k_2.k_1\right){}^2 k_1.\epsilon _2.k_1 k_1.\epsilon _3.\epsilon _4.k_3+64 k_2.k_1 k_3.k_1 k_1.\epsilon _2.k_1 k_1.\epsilon _3.\epsilon _4.k_3\nonumber\\&&+32 \left(k_3.k_1\right){}^2 k_1.\epsilon _2.k_1 k_1.\epsilon _3.\epsilon _4.k_3+64 \left(k_2.k_1\right){}^2 k_1.\epsilon _2.k_3 k_1.\epsilon _3.\epsilon _4.k_3\nonumber\\&&+64 k_2.k_1 k_3.k_1 k_1.\epsilon _2.k_3 k_1.\epsilon _3.\epsilon _4.k_3+32 \left(k_2.k_1\right){}^2 k_3.\epsilon _2.k_3 k_1.\epsilon _3.\epsilon _4.k_3\nonumber\\&&+32 k_2.k_1 k_3.k_1 k_1.\epsilon _2.k_3 k_2.\epsilon _3.\epsilon _4.k_2+32 k_2.k_1 k_3.k_1 k_1.\epsilon _2.k_1 k_2.\epsilon _3.\epsilon _4.k_3\nonumber\\&&+32 \left(k_3.k_1\right){}^2 k_1.\epsilon _2.k_1 k_2.\epsilon _3.\epsilon _4.k_3+32 k_2.k_1 k_3.k_1 k_1.\epsilon _2.k_3 k_2.\epsilon _3.\epsilon _4.k_3\nonumber\\&&-32 \left(k_2.k_1\right){}^2 k_2.\epsilon _4.k_3 k_3.\epsilon _2.\epsilon _3.k_1-32 \left(k_2.k_1\right){}^2 k_1.\epsilon _3.k_2 k_3.\epsilon _2.\epsilon _4.k_2\nonumber\\&&-32 \left(k_2.k_1\right){}^2 k_1.\epsilon _3.k_2 k_3.\epsilon _2.\epsilon _4.k_3-32 \left(k_2.k_1\right){}^2 k_3.k_1 k_1.\epsilon _2.\epsilon _3.\epsilon _4.k_3\nonumber\\&&-32 k_2.k_1 \left(k_3.k_1\right){}^2 k_1.\epsilon _2.\epsilon _3.\epsilon _4.k_3+32 \left(k_2.k_1\right){}^3 k_1.\epsilon _2.\epsilon _4.\epsilon _3.k_1\nonumber\\&&+64 \left(k_2.k_1\right){}^2 k_3.k_1 k_1.\epsilon _2.\epsilon _4.\epsilon _3.k_1+32 k_2.k_1 \left(k_3.k_1\right){}^2 k_1.\epsilon _2.\epsilon _4.\epsilon _3.k_1\nonumber\\&&+32 \left(k_2.k_1\right){}^2 k_3.k_1 k_1.\epsilon _2.\epsilon _4.\epsilon _3.k_2+32 k_2.k_1 \left(k_3.k_1\right){}^2 k_1.\epsilon _2.\epsilon _4.\epsilon _3.k_2\nonumber\\&&-32 \left(k_2.k_1\right){}^3 k_1.\epsilon _3.\epsilon _2.\epsilon _4.k_2-32 \left(k_2.k_1\right){}^2 k_3.k_1 k_1.\epsilon _3.\epsilon _2.\epsilon _4.k_2\nonumber\\&&-32 \left(k_2.k_1\right){}^2 k_3.k_1 k_2.\epsilon _3.\epsilon _2.\epsilon _4.k_2-32 \left(k_2.k_1\right){}^2 k_3.k_1 k_2.\epsilon _3.\epsilon _2.\epsilon _4.k_3\nonumber\\&&-32 \left(k_2.k_1\right){}^2 k_3.k_1 k_3.\epsilon _2.\epsilon _3.\epsilon _4.k_2-32 \left(k_2.k_1\right){}^2 k_3.k_1 k_3.\epsilon _2.\epsilon _3.\epsilon _4.k_3\nonumber\\&&+32 \left(k_2.k_1\right){}^3 k_3.\epsilon _2.\epsilon _4.\epsilon _3.k_1+32 \left(k_2.k_1\right){}^2 k_3.k_1 k_3.\epsilon _2.\epsilon _4.\epsilon _3.k_1\nonumber\\&& -32 \left(k_2.k_1\right){}^3 k_3.k_1 {\Tr}\left[\epsilon _2.\epsilon _3.\epsilon _4\right]-32 \left(k_2.k_1\right){}^2 \left(k_3.k_1\right){}^2  {\Tr}\left[\epsilon _2.\epsilon _3.\epsilon _4\right]\bigg]\phi_1\labell{pr}
\eeqa
where $\eps_2$ is the polarization of the graviton. We have also divided the amplitude by $4\sqrt{2}$ to be consistent with the normalization of fields in sec.1.

The above amplitude should be compared with the corresponding couplings in \reef{F1F3}. 
Using the expansion
\beqa
\cH^T_{h q r,n}\cM_{,rm}\cH_{k n p,h} &=& -e^{-\phi}\phi_{,rm}H_{h q r,n}H_{k n p,h}+\cdots
\eeqa
where dots refer to the other terms which involve the R-R fields, one finds the following couplings at the tree-level in the Einstein frame:
\beqa
S &\supset&\frac{\gamma\z(3)}{8 } \int d^{10}x\sqrt{-G}e^{-5\phi/2}\bigg[ -8 \phi_{,rm} H_{h q r,n} H_{k n p,h} R_{m p k q}+8 \phi_{,mk} H_{n p r,h} H_{h n q,r} R_{k q m p}\nonumber\\&&+8 \phi_{,qh} H_{n p q,m} H_{m p r,k} R_{k r h n}-8 \phi_{,qm} H_{n p q,h} H_{k n r,h} R_{m r k p}+4 \phi_{,mh} H_{n p q,h} H_{n p r,k} R_{m r k q}\nonumber\\&&-4\phi_{,rk} H_{m n q,h}  H_{n p q,k} R_{p r h m}+4 \phi_{,rm}H_{m n p,k}  H_{n p q,h} R_{q r h k}\bigg]\labell{F1F31}
\eeqa
In the string frame,   the overall dilaton factor  becomes $e^{-2\phi}$. We have checked explicitly that the momentum space of the above couplings   are exactly the same as the couplings in \reef{pr}.

\section{Five-point  amplitude}

The coupling $\eps_{10}\eps_{10}R^4$ in \reef{Y1} is  a total derivative term at four graviton level \cite{Zumino:1985dp}, so its presence in the effective action can not be confirmed by the S-matrix element of four gravitons.  The $\sigma$-model beta function approach implies the presence of this term at the tree-level \cite{Grisaru:1986vi,Freeman:1986zh}. To 
confirm    this term directly from the sphere-level S-matrix calculations, one needs to study the scattering amplitude of five gravitons.   
In this section we are going to show that the sphere-level scattering amplitude of five gravitons confirms the presence of the coupling $\eps_{10}\eps_{10}R^4$ at the tree-level. The  one-loop calculation which confirms   the presence of this term in \reef{Y2}, appears in \cite{Richards:2008jg,Liu:2013dna}
 
To confirm the tree-level couplings \reef{Y1} by the sphere-level S-matrix element of five graviton vertex operators, one has to first calculate explicitly the latter amplitude and then expand it at  low energy to find massless poles and contact terms. The massless poles and the contact terms of the string theory amplitude should be the same as the massless poles and the contact terms  of the corresponding  Feynman amplitude in the field theory. In the field theory side, the eight-derivative massless poles are  produced  by the Hilbert-Einstein term in the supergravity \reef{super} and  by the four-graviton couplings in $t_8t_8R^4$. The eight-derivative contact terms of the   S-matrix element should also be the same as the five-graviton couplings in \reef{Y1}.  

We are not going to calculate the S-matrix element explicitly  in the string theory side, however, we use the observation made in \cite{Garousi:2012yr,Barreiro:2012aw} that indicates the sphere-level S-matrix elements should have no term  with zero or one Mandelstam variable $k_i\inn k_j$. In other words, the terms in the S-matrix elements should have at least two Mandelstam variables. This is resulted from the fact that in the disk-level scattering amplitude of open string gauge bosons, two of the vertex operators must be in $(-1)$-picture and all other must be in $0$-picture. As a result, the open string amplitude  has no term with zero  Mandelstam variable. Using the KLT prescription, one finds that the closed string amplitudes at sphere-level have at least two Mandelstam variables.

The scattering amplitude of five on-shell gravitons in field theory side is given by the following Feynman amplitude:
\beqa
A &=&\sum_{i=1}^{10}A_i  +V_{R^4}(1,2,3,4,5)\labell{A}
\eeqa
where the massless poles $A_i$ are given as
\beqa
A_1&=&V_{R^4}(1,2,3,h)_{mn}G_{mn,hk}V_{R}(h,4,5)_{hk}\nonumber\\
A_2&=&V_{R^4}(1,2,4,h)_{mn}G_{mn,hk}V_{R}(h,3,5)_{hk}\nonumber\\
A_3&=&V_{R^4}(1,4,3,h)_{mn}G_{mn,hk}V_{R}(h,2,5)_{hk}\nonumber\\
A_4&=&V_{R^4}(4,2,3,h)_{mn}G_{mn,hk}V_{R}(h,1,5)_{hk}\nonumber\\
A_5&=&V_{R^4}(1,2,5,h)_{mn}G_{mn,hk}V_{R}(h,3,4)_{hk}\nonumber\\
A_6&=&V_{R^4}(1,3,5,h)_{mn}G_{mn,hk}V_{R}(h,2,4)_{hk}\nonumber\\
A_7&=&V_{R^4}(3,2,5,h)_{mn}G_{mn,hk}V_{R}(h,1,4)_{hk}\nonumber\\
A_8&=&V_{R^4}(1,4,5,h)_{mn}G_{mn,hk}V_{R}(h,2,3)_{hk}\nonumber\\
A_9&=&V_{R^4}(2,4,5,h)_{mn}G_{mn,hk}V_{R}(h,1,3)_{hk}\nonumber\\
A_{10}&=&V_{R^4}(3,4,5,h)_{mn}G_{mn,hk}V_{R}(h,1,2)_{hk}\nonumber\\
\eeqa
Our notation for the vertexes is that $h$ appears for the off-shell graviton    whereas  the particle number appears for  the on-shell graviton. For example, in the vertex $V_{R}(h,4,5)_{hk}$ one of the gravitons is off-shell and the other two gravitons are on-shell with particle labels $4$ and $5$. The propagator and the vertex $V_{R}(h,4,5)_{hk}$ are read from the supergravity \reef{super}
\beqa
G_{mn,hk}&\!\!\!\!\!\!\!\!=\!\!\!\!\!\!\!\!&-\frac{i}{2k.k}\left(\eta _{h m}\eta _{k n}+\eta _{h n}\eta _{k m}-\frac{1}{4}\eta _{h k}\eta _{m n}\right)\labell{pro}\\
V_{R}(h,4,5)_{hk}&\!\!\!\!\!\!\!\!\!\!\!\!=\!\!\!\!\!\!\!\!\!\!\!\!&-2i\bigg[-\frac{1}{4} k_4.\varphi _5 k_5.\varphi _4 \rho _4.\rho _5 \eta _{h k}-\frac{1}{4} k_4.\rho _5 k_5.\varphi _4 \rho _4.\varphi _5 \eta _{h k}-\frac{1}{4} k_4.\varphi _5 k_5.\rho _4 \varphi _4.\rho _5 \eta _{h k}\nonumber\\&&+\frac{3}{4} k_4.k_5 \rho _4.\varphi _5 \varphi _4.\rho _5 \eta _{h k}-\frac{1}{4} k_4.\rho _5 k_5.\rho _4 \varphi _4.\varphi _5 \eta _{h k}+\frac{3}{4} k_4.k_5 \rho _4.\rho _5 \varphi _4.\varphi _5 \eta _{h k}\nonumber\\&&-\frac{1}{2} \rho _4.\varphi _5 \varphi _4.\rho _5 \left(k_4\right)_h \left(k_4\right)_k-\frac{1}{2} \rho _4.\rho _5 \varphi _4.\varphi _5 \left(k_4\right)_h \left(k_4\right)_k-\frac{1}{2} \rho _4.\varphi _5 \varphi _4.\rho _5 \left(k_4\right)_k \left(k_5\right)_h\nonumber\\&&-\frac{1}{2} \rho _4.\rho _5 \varphi _4.\varphi _5 \left(k_4\right)_k \left(k_5\right)_h-\frac{1}{2} \rho _4.\varphi _5 \varphi _4.\rho _5 \left(k_5\right)_h \left(k_5\right)_k-\frac{1}{2} \rho _4.\rho _5 \varphi _4.\varphi _5 \left(k_5\right)_h \left(k_5\right)_k\nonumber\\&&+\frac{1}{2} k_4.\varphi _5 \varphi _4.\rho _5 \left(k_4\right)_k \left(\rho _4\right)_h+\frac{1}{2} k_4.\rho _5 \varphi _4.\varphi _5 \left(k_4\right)_k \left(\rho _4\right)_h+\frac{1}{2} k_5.\varphi _4 \rho _4.\varphi _5 \left(k_5\right)_k \left(\rho _5\right)_h\nonumber\\&&+\frac{1}{2} k_5.\rho _4 \varphi _4.\varphi _5 \left(k_5\right)_k \left(\rho _5\right)_h+\frac{1}{2} k_4.\varphi _5 k_5.\varphi _4 \left(\rho _4\right)_k \left(\rho _5\right)_h-\frac{1}{2} k_4.k_5 \varphi _4.\varphi _5 \left(\rho _4\right)_k \left(\rho _5\right)_h\nonumber\\&&+\frac{1}{2} k_4.\varphi _5 \rho _4.\rho _5 \left(k_4\right)_k \left(\varphi _4\right)_h+\frac{1}{2} k_4.\rho _5 \rho _4.\varphi _5 \left(k_4\right)_k \left(\varphi _4\right)_h-k_4.\rho _5 k_4.\varphi _5 \left(\rho _4\right)_k \left(\varphi _4\right)_h\nonumber\\&&+\frac{1}{2} k_4.\varphi _5 k_5.\rho _4 \left(\rho _5\right)_k \left(\varphi _4\right)_h-\frac{1}{2} k_4.k_5 \rho _4.\varphi _5 \left(\rho _5\right)_k \left(\varphi _4\right)_h+\frac{1}{2} k_5.\varphi _4 \rho _4.\rho _5 \left(k_5\right)_k \left(\varphi _5\right)_h\nonumber\\&&+\frac{1}{2} k_5.\rho _4 \varphi _4.\rho _5 \left(k_5\right)_k \left(\varphi _5\right)_h+\frac{1}{2} k_4.\rho _5 k_5.\varphi _4 \left(\rho _4\right)_k \left(\varphi _5\right)_h-\frac{1}{2} k_4.k_5 \varphi _4.\rho _5 \left(\rho _4\right)_k \left(\varphi _5\right)_h\nonumber\\&&-k_5.\rho _4 k_5.\varphi _4 \left(\rho _5\right)_k \left(\varphi _5\right)_h+\frac{1}{2} k_4.\rho _5 k_5.\rho _4 \left(\varphi _4\right)_k \left(\varphi _5\right)_h-\frac{1}{2} k_4.k_5 \rho _4.\rho _5 \left(\varphi _4\right)_k \left(\varphi _5\right)_h\bigg]\nonumber
\eeqa
where we have written the graviton polarization as 
\beqa
\varepsilon_{hk}&=&\frac{1}{2}(\rho_h\varphi_k+\rho_k\varphi_h)
\eeqa
to impose the symmetry of the graviton. The vertex $ V_{R^4}(1,2,3,h)_{mn}$ should be read from the action \reef{Y1}. Since $\eps_{10}\eps_{10}R^4$ is a total derivative at four graviton level, this vertex must be read from $t_8t_8R^4$. 

 The $t_8$ tensor was first defined in \cite{Schwarz:1982jn}, \ie the contraction of $t_8$ with   four arbitrary antisymmetric matrices $M^1,\,\cdots, M^4$  is  
\beqa
&&t_{hkmnpqrs}M^1_{hk}M^2_{mn}M^3_{pq}M^4_{rs}=8(\tr M^1M^2M^3M^4+\tr M^1M^3M^2M^4+\tr M^1M^3M^4M^2)\nonumber\\
&&\qquad\qquad\qquad\quad-2(\tr M^1M^2\tr M^3M^4+\tr M^1M^3\tr M^2M^4+\tr M^1M^4\tr M^2M^3)\labell{t8}
\eeqa
Using this tensor, the coupling  $t_8t_8R^4$ can be written as \cite{Gross:1986mw}\footnote{For the generalized Riemann curvature \reef{trans} which does not have all symmetries of the Riemann curvature, the coupling  $t_8t_8\bar{R}^4$ has different structure \cite{Garousi:2013zca}.}
\beqa
t_8t_8R^4&\equiv&t_{\mu_1\cdots \mu_8}t_{\nu_1\cdots \nu_8}R_{\mu_1\mu_2}{}_{\nu_1\nu_2}R_{\mu_3\mu_4}{}_{\nu_3\nu_4}R_{\mu_5\mu_6}{}_{\nu_5\nu_6}R_{\mu_7\mu_8}{}_{\nu_7\nu_8}\nonumber\\
&=&3\cdot 2^7\bigg[R_{hkmn}R_{krnp}R_{rs}{}_{qm}R_{sh}{}_{pq}+\frac{1}{2}R_{hkmn}R_{krnp}R_{rspq}R_{shqm}\labell{Y3}\\
&&  -\frac{1}{2}R_{hkmn}R_{krmn}R_{rspq}R_{shpq} -\frac{1}{4}R_{hkmn}R_{krpq}R_{rs}{}_{mn}R_{sh}{}_{pq}\nonumber\\
&& +\frac{1}{16}R_{hkmn}R_{khpq}R_{rsmn}R_{srpq}+\frac{1}{32}R_{hkmn}R_{khmn}R_{rspq}R_{srpq}\bigg]\nonumber
\eeqa
Using the linearized Riemann curvature \reef{R}, one finds the vertex 
\beqa
 V_{R^4}(1,2,3,h)_{mn}&=&3\cdot 2^7i\gamma\z(3)\bigg[k_1.\varphi _2 k_1.\varphi _3 k_2.\rho _3 k_2.\varphi _1 k_3.\rho _1 k_3.\rho _2 \left(k_1\right)_m \left(k_1\right)_n+ \cdots\bigg]\labell{R4}
 \eeqa
 where dots refer to all other terms which are too many to be able to write them. In the supergravity vertex \reef{pro}, the terms which are proportional to $k_4\inn k_5$ produce contact terms  when they are multiplied by the propagator. We call them $A_i^c$. The other terms produce massless poles in \reef{A}. We call them $A_i^{\rm pole}$. So the massless poles in \reef{A} are
 \beqa
 A^{\rm pole}&=&\sum_{i=1}^{10}A^{\rm pole}_i
 \eeqa
 We have checked that these massless poles are proportional to  at least two Mandelstam variables, as expected from the corresponding sphere-level scattering amplitude in the string theory side \cite{Garousi:2012yr,Barreiro:2012aw}. 
 
 On the other hand, the contact terms $A_i^c$ do not satisfy this constraint, \ie they have terms with no Mandelstam variable. These contact terms however must be combined with the contact term $V_{R^4}(1,2,3,4,5)$ in \reef{A} to satisfy this constraint. 
 So we expect the following contact terms:
\beqa
A^c &=&\sum_{i=1}^{10}A^c_i  +V_{R^4}(1,2,3,4,5)\labell{Ac}
\eeqa 
to be proportional to  at least two Mandelstam variables. The five   graviton  couplings of $t_8t_8R^4$ are resulted from two sources. One source is to use \reef{R} for all  Riemann curvatures   and the expansions $G^{ab}= -2h^{ab} $ for one metric and $G^{ab}=\eta^{ab}$ for all the other  metrics in the contracted indices. The other source is to use \reef{R} for three Riemann curvatures, use the second order gravitons  for one of the Riemann curvature, \ie
\beqa
R_{ a m n b}&=&-h_{b m,t} h_{a n,t}+h_{a b,t} h_{m n,t}-h_{m n,t} h_{a t,b}+h_{b m,t} h_{a t,n}-h_{m n,t} h_{b t,a}+h_{a n,t} h_{b t,m}\nonumber\\&&-h_{a t,n} h_{b t,m}+h_{a n,t} h_{m t,b}-h_{a t,n} h_{m t,b}-h_{a b,t} h_{m t,n}+h_{a t,b} h_{m t,n}+h_{b t,a} h_{m t,n}\nonumber\\&&+h_{b m,t} h_{n t,a}-h_{b t,m} h_{n t,a}-h_{m t,b} h_{n t,a}-h_{a b,t} h_{n t,m}+h_{a t,b} h_{n t,m}+h_{b t,a} h_{n t,m}\labell{R2}
\eeqa 
and to use the expansion $G^{ab}=\eta^{ab}$ for all the metrics in the contracted indices.  
 Using the above two sources to calculate the contact terms in $V_{R^4}(1,2,3,4,5)$, one still finds  that the amplitude \reef{Ac} has terms with no Mandelstam variable. So there must be another term beside the coupling $t_8t_8R^4$.

Now consider the coupling $\eps_{10}\eps_{10}R^4$. Using the relation $\frac{1}{2}\eps_{\mu\nu\mu_1\cdots \mu_8}\eps^{\mu\nu\nu_1\cdots\nu_8}=-8!\delta^{\nu_1}_{[\mu_1}\cdots \delta^{\nu_8}_{\mu_8]}$, one may write 
\beqa
\frac{1}{2}\eps_{10}\eps_{10}R^4&=&3\cdot 2^9\bigg[-R_{h k m n} R_{p q k r} R_{n r q s} R_{m s h p}+R_{h k m n} R_{p q h k} R_{n r q s} R_{m s p r}\nonumber\\&&+\frac{1}{2} R_{h k m n} R_{p q r s} R_{m r p q} R_{n s h k}-\frac{1}{2} R_{h k m n} R_{p q r s} R_{m r h p} R_{n s k q}\nonumber\\&&-\frac{1}{16} R_{h k m n} R_{m n p q} R_{p q r s} R_{r s h k}-\frac{1}{32} R_{h k m n} R_{m n h k} R_{p q r s} R_{r s p q}\nonumber\\&&-2 R_{h  m} R_{n p q r} R_{q r n s} R_{m s h p}+R_{h  m} R_{n p q r} R_{q r h s} R_{m s n p}+4 R_{h  m} R_{m n p q} R_{q r h s} R_{p s n r}\nonumber\\&&+\frac{1}{3} R R_{m n p q} R_{q r n s} R_{p s m r}-\frac{1}{12} R R_{m n p q} R_{p q r s} R_{r s m n}+\cdots\bigg]
\eeqa
where dots refer to the terms with more than one Ricci or scalar curvature. Since we are interested in the couplings of five on-shell gravitons in $V_{R^4}(1,2,3,4,5)$, such terms have no contribution. The Riemann curvature couplings in the first three lines above have two sources in producing the five graviton contact terms, as we have explained in the previous paragraph. The Ricci curvature terms in the fourth line above have also two sources. One source is to replace the Riemann curvature terms with the linear form \reef{R}, the Ricci curvature with the nonlinear form \reef{R2} and    $G^{ab}=\eta^{ab}$ for all the metrics in the contracted indices. The second source is to replace the Ricci curvature with $R_{ab}\equiv R_{acbc}=-2h^{cd}R_{acbd}$, and then to replace all the Riemann curvatures with the linear form \reef{R} and all the contracted metrics with $G^{ab}=\eta^{ab}$. The scalar curvature couplings in the last line above have one source in producing the five on-shell graviton contact terms. The scalar curvature must be replaced by the nonlinear form \reef{R2}, the Riemann curvatures with the linear form \reef{R} and all the contracted metrics must be replaced by  $G^{ab}=\eta^{ab}$.

Calculating the contact term  $V_{R^4}(1,2,3,4,5)$ from the couplings $t_8t_8R^4+\frac{\alpha}{2}\eps_{10}\eps_{10}R^4$ as we have explained above, and using the conservation of momentum and the on-shell relations $k_i\inn k_i=k_i\inn\veps_i=\veps_i\inn k_i=0$ to write the amplitude in terms of the independent Mandelstam variables $k_1\inn k_2,\, k_1\inn k_3,\, k_2\inn k_3,\, k_2\inn k_4,\, k_3\inn k_4$ and in terms of independent variables $k_1\inn \veps_2,\,k_1\inn \veps_3,\,k_1\inn \veps_4$, $k_2\inn \veps_1,\,k_2\inn \veps_3,\,k_2\inn \veps_4,\,k_2\inn \veps_5$,  $k_3\inn \veps_1,\,k_3\inn \veps_2,\,k_3\inn \veps_4,\,k_3\inn \veps_5$,  $k_4\inn \veps_1,\,k_4\inn \veps_2,\,k_4\inn \veps_3,\,k_4\inn \veps_5$, we have found that the contact terms \reef{Ac} have at least two Mandelstam variables when the constant $\alpha=\frac{1}{4}$. This confirms that the tree-level couplings in \reef{Y1} are consistent with sphere-level scattering amplitude of five gravitons in the RNS formalism. As a further check, one may calculate explicitly the scattering amplitude in the string theory side which contains only terms proportional to at least two Mandelstam variables,  and compare  it with the field theory \reef{Y1}.

Finally, let us mention that the coupling $t_8t_8R^4$ is invariant under standard  T-duality \cite{Garousi:2012jp,Garousi:2013zca,Liu:2013dna}, whereas the coupling $\eps_{10}\eps_{10}R^4$ is not invariant under the standard linear T-duality in the absence of B-field . To see this, one may consider the dimensional reduction of this coupling on a circle and may consider the following term
\beqa
 \eps_{\mu\nu y\mu_2\cdots\mu_8}\eps_{\mu\nu y\nu_2\cdots\nu_8}R_{y\mu_2}{}_{y\nu_2}R_{\mu_3\mu_4}{}_{\nu_3\nu_4}R_{\mu_5\mu_6}{}_{\nu_5\nu_6}R_{\mu_7\mu_8}{}_{\nu_7\nu_8}
\eeqa
where $y$ is the Killing coordinate. Under the standard linear T-duality $R_{y\mu_2}{}_{y\nu_2}$ goes to  $-R_{y\mu_2}{}_{y\nu_2}$ \cite{Garousi:2012yr}. So the coupling $\eps_{10}\eps_{10}R^4$ is not invariant under the linear T-duality. However, this term is a total derivative at four  field level so it has no effect in the action at four field level. This indicates that only the action, not the Lagrangian density,  must be invariant under T-duality. Moreover, using the fact that the Gauss-Bonnet term $\eps_{10}\eps_{10}R^4$ is the only coupling which is total derivative at four field level and the observation   that the couplings in  $t_8t_8R^4$ are the most general  couplings which are consistent with the standard linear T-duality  \cite{Garousi:2013zca}, one concludes that there is no other gravity term in \reef{Y1} at order $\alpha'^3$.

The coupling $\eps_{10}\eps_{10}R^4$,  however, must be extended to T-duality invariant form because this term has significant effect at higher levels. In the absence of the B-field, the only way to make it   invariant under the standard T-duality is to include couplings which involve dilaton in the string frame. It has been argued in \cite{Liu:2013dna} that there is no dilaton couplings in the string frame. However, the argument in \cite{Liu:2013dna} is based on the specific form of the effective action of the heterotic  string theory at order $\alpha'$ which has no dilaton in the string frame. This action is related to the standard form of the heterotic effective action  which is manifestly T-duality invariant and includes various dilaton couplings \cite{Kaloper:1997ux}, by some field redefinitions. The field redefinition, however, changes the standard T-duality transformations   to non-standard forms. Therefore,  we expect the couplings like 
\beqa
 \eps_{\mu\nu \rho\mu_2\cdots\mu_8}\eps_{\mu\nu \rho\nu_2\cdots\nu_8}\prt_{\mu_2}\phi\prt_{\nu_2}\phi R_{\mu_3\mu_4}{}_{\nu_3\nu_4}R_{\mu_5\mu_6}{}_{\nu_5\nu_6}R_{\mu_7\mu_8}{}_{\nu_7\nu_8}
\eeqa
 to be included in the type II theory in the string frame, to make the whole action to be invariant under the standard T-duality transformations. It would be interesting to find such dilaton couplings.

{\bf Acknowledgments}:    This work is supported by Ferdowsi University of Mashhad under grant 2/20625.

\end{document}